\documentstyle[12pt]{article}
\begin{document}

\newcommand{\ep}{\varepsilon}
\newcommand{\fii}{\varphi}
\newcommand{\la}{\lambda}

\setcounter{page}{0}
\thispagestyle{empty}

{}~\vspace{2cm}

\begin{center} {\Large {\bf
Compact analytical form for a class of three-loop vacuum Feynman diagrams
}} \\
 \vspace{1.5cm}
 {\large
  A.V.Kotikov
\\
 \vspace {0.5cm}
 {\em Particle Physics Laboratory\\
  JINR, Dubna, 141980 Russia.\\ e-mail: kotikov@sunse.jinr.ru}}
\end{center}

\vspace{1.5cm}\noindent
\begin{center} {\bf Abstract} \end{center}

We present compact, fully analytical expressions for singular parts of a 
class of three-loop diagrams which cannot be factorized into lower-loop 
integrals. As a result of the calculations we obtain the analytical 
expression for the three-loop effective potential of the massive $O(N)$
$\varphi ^4$ model presented recently by J.-M.Chung and B.K.Chung, Phys.
Rev.D{\bf 56}, 6508 (1997).


\pagestyle{plain}


Though Quantum Field Theory
 already has a long history and a
number of different approaches, Feynman diagrams (FD) are still the
main source of
its dynamical information.
Vacuum (bubble) FD (without external momenta)
considered here
have several points of 
applications.
First of all, it is the evaluation of the effective potentials
(for the recent study see \cite{ChCh} and references therein) and
renormalization group characteristics ($\beta$-functions and anomalous 
dimensions) of  quantum field
models and specific operators (for example, 
anomalous dimensions of operators in the Wilson expansion).
The second important place of the applicability is the calculations of
various
processes (essentially in Standard Model), where there is a possibility to 
neglect most of the masses and momenta and to calculate only some
Taylor coefficients of multipoint FD. These coefficients are bubbles
having different (sometimes quite big) powers of propagators
(i.e. indices of propagators). Using different recurrence relations 
\cite{Bro}-\cite{Avdeev} it is possible usually to represent these Taylor 
coefficients as sets of very simple (usually one-loop) bubbles and several
so-called
master integrals, which cannot be factorized into sums of lower-loop 
integrals. These recurrence relations are some particial cases of the relation
\cite{DEM,DEMN} for a general $n$-point (sub)graph with masses of its lines
$m_1, m_2, ..., m_n$, line momenta $p_1, p_2=p_1 - p_{12}, p_n=p_1 - p_{1n}$
and indeces $j_1, j_2, .., j_n$, respectively,
\begin{eqnarray}
0&=& \int d^Dp_1 \frac{\partial}{\partial p_1^{\mu}} {\biggl(\prod_{i=1}^{n}
c_i^{j_i} \biggr)}^{-1} \label{1} \\ 
&=& \int d^Dp_1  {\biggl(\prod_{i=1}^{n}
c_i^{j_i} \biggr)}^{-1} \Biggl[ D - 2j_1 \biggl( 1- \frac{m_1^2}{c_1} \biggr) 
- \sum_{i=2}^{n} j_i 
 \biggl( 1-
\frac{m_1^2 + m_i^2 + p^2_{1i}-c_1}{c_i} \biggr)
\Biggr],
\nonumber
\end{eqnarray}
where
$c_k=p^2_k+m^2_k$ are the propagators of $n$-point (sub)graph. The
equation (\ref{1}) is based on the rule of integration by part \cite{VPH}
and has been discovered in the process of the calculation of propagator- 
and vertex-type diagrams (in \cite{DEM} for the case $n=3$) and of $n$-point
diagrams (in \cite{DEMN} for the arbitrary $n$).

Because the diagram with the index $(i+1)$ of the propagator $c_i$ may be 
represented as the derivative (on the mass $m_i$), Eq.(\ref{1}) leads to 
the differential equations (in principle, to partial differential equations)
for the initial diagram (having the index $i$, respectively). This approach
which is based on
 the Eq.(\ref{1}) and allows to construct the (differential) relations
between diagrams has been named as Differential Equations Method (DEM).  
 For most
interested cases (where the number of the masses is limited)
these partial differential equations may be represented through
original differential equation\footnote{The example of the direct application
of the partial differential equation may be found in \cite{FJ1}.},
which is usually simpler to analyse. For the bubbles, DEM has been 
applied in \cite{FJ,PV}.

In the case of large number of diagrams (for example, in the calculation
of various processes of Standard Model)
it is convenient \cite{FT, Avdeev}
to use Eq.(\ref{1}) without representating the diagrams in its r.h.s. as 
derivatives of initial ones. Then Eq.(\ref{1}) gives the connections between
different diagrams decreasing essentially the number of
complicated integrals. These integrals (i.e. master diagrams)
may be calculated \cite{FKV}
with help of DEM as it was discussed above.\\

{\bf 1.} In the present article we consider a class of three-loop bubbles (see
Fig. 1) having two different masses: the masses of thick lines are one-third
of the masses of the bold lines. The diagrams are interested 
because they cannot be factorized into lower-loop integrals (see discussions
in \cite{ChCh1}).
Moreover, some of them ($J(a)$, $K(a)$, $L(a)$ and $M(a)$) give contributions
(see \cite{ChCh,ChCh1}) to the three-loop effective potential of the massive 
$O(N)$ $\varphi ^4$-model. The diagrams have been studied recently in 
\cite{ChCh1} using the methods developed in the articles
\cite{CMM,Kast}\footnote{The singular part of $M(a)$, $M(b)$ and $M(c)$
diagrams has been calculated in \cite{PV} using DEM \cite{DEM}. The
regular part of $M(a)$ has been found very recently by Broadhurst
\cite{Brobu}.}.
The following results for their singular parts have been
found\footnote{Contrary to \cite{ChCh1} we use the
space $D=4-2\ep$ and sum the terms
$\gamma ^n$
($\gamma$ - is Euler constant) and $\zeta_2^n$ to exponents.}:
\begin{eqnarray}
J(a) &=& N_2 \biggl[\frac{2}{\ep^3} +
\frac{23}{3} \frac{1}{\ep^2} +\frac{35}{2} \frac{1}{\ep} 
+ O(1) \biggr] \nonumber \\
J(b) &=& N_2 \biggl[\frac{22}{27}\frac{1}{\ep^3} +
\frac{1}{\ep^2}\Bigl( \frac{83}{27} + \frac{7}{9}\log {3} \Bigr)
+ \frac{1}{\ep}\Bigl( \frac{365}{54} + \frac{55}{18}\log {3} +
\frac{1}{2}\log^2 {3} 
\Bigr) + O(1) \biggr] \nonumber \\
J(c) &=& N_2 \biggl[\frac{2}{9}\frac{1}{\ep^3} +
\frac{1}{\ep^2}\Bigl( \frac{23}{27} + \frac{2}{3}\log {3} \Bigr)
+ \frac{1}{\ep}\Bigl( \frac{35}{18} + \frac{23}{9}\log {3} + \log^2 {3} 
\Bigr) + O(1) \biggr] \nonumber \\
K(a) &=& -N_1 \biggl[\frac{1}{\ep^3} +
\frac{17}{3} \frac{1}{\ep^2} + \frac{1}{\ep}\Bigl( \frac{67}{3} + 6 A \Bigr) 
+ O(1) \biggr] \nonumber \\
K(b) &=& -N_1 \biggl[\frac{7}{9}\frac{1}{\ep^3} +
\frac{1}{\ep^2}\Bigl( \frac{13}{3} + \frac{1}{3}\log {3} \Bigr)
+ \frac{1}{\ep}\Bigl( \frac{151}{9} + 3 A +\frac{1}{3}B +2\log {3} 
\Bigr) + O(1) \biggr] \nonumber \\
K(c) &=& -N_1 \biggl[\frac{5}{9}\frac{1}{\ep^3} +
\frac{1}{\ep^2}\Bigl( \frac{29}{9} + \frac{2}{3}\log {3} \Bigr)
+ \frac{1}{\ep}\Bigl( \frac{37}{3} + \frac{2}{3}B +
4\log {3}  
\Bigr) + O(1) \biggr] \label{2} \\
K(d) &=& -N_1 \biggl[\frac{5}{9}\frac{1}{\ep^3} +
\frac{1}{\ep^2}\Bigl( 3 + \frac{2}{3}\log {3} \Bigr)
+ \frac{1}{\ep}\Bigl( \frac{101}{9} + \frac{2}{3}B +
4\log {3}  
\Bigr) + O(1) \biggr] \nonumber \\
L(a) &=& N_0 \biggl[\frac{1}{3} \frac{1}{\ep^3} +
\frac{2}{3} \frac{1}{\ep^2} + \frac{1}{\ep}\Bigl( \frac{2}{3} + 2 A \Bigr) 
+ O(1) \biggr] \nonumber \\
L(b) &=& N_0 \biggl[\frac{1}{3} \frac{1}{\ep^3} +
\frac{2}{3} \frac{1}{\ep^2} + \frac{1}{\ep}\Bigl( \frac{2}{3} + A +C \Bigr) 
+ O(1) \biggr] \nonumber \\
L(c) &=& N_0 \biggl[\frac{1}{3} \frac{1}{\ep^3} +
 \frac{1}{\ep^2} \Bigl( \frac{2}{3} + \log {3} \Bigr)
+ \frac{1}{\ep}\Bigl( \frac{2}{3} + 2 B
+2\log {3} +  \frac{3}{2} \log^2 {3} \Bigr) 
+ O(1) \biggr] \nonumber \\
L(d) &=& N_0 \biggl[\frac{1}{3} \frac{1}{\ep^3} +
\frac{2}{3} \frac{1}{\ep^2} + \frac{1}{\ep}\Bigl( \frac{2}{3} + 2 C \Bigr) 
+ O(1) \biggr] \nonumber \\
M(a) &=& M(b) = M(c) = N_0 \biggl[\frac{2}{\ep} \zeta_3
+ O(1) \biggr], \nonumber
\nonumber
\end{eqnarray}
where
the normaliation factor 
$$
N_k = \frac{m^{2k}}{(4\pi)^6} {(\frac{m^2}{\overline \mu^2})}^{-3\ep}
\cdot \exp{(3/2 \zeta_2 \ep^2)}
~~~ \mbox{ and } ~~~ \overline \mu^2 = (4\pi \mu^2)e^{\gamma}  
$$
and $\zeta_n$ are Euler $\zeta$-functions.

The constants $A$, $B$ and $C$ have been represented \cite{ChCh1} in the form:
\begin{eqnarray}
A=f(1,1),~~~B=f(1,3)~~~\mbox{ and }~~~C=f(\frac{1}{3},\frac{1}{3}),
\nonumber
\end{eqnarray}
where
\begin{eqnarray}
f(a,b) = \int_0^1 dx \biggl[ \int^{1-z}_0 dy \biggl(-\frac{\log{(1-y)}}{y}
\biggr)
-\frac{z\log{z}}{1-z} \biggr],~~~z=\frac{ax+b(1-x)}{x(1-x)}
\label{3}
\end{eqnarray}

The puprose of this short letter is to calculate analytically
$A$, $B$ and $C$ constants and, thus, to obtain exact results for the singular
parts of the FD presented in Fig. 1.

Before evaluations we would like to stress that the knowledge of
the exact values for the singular parts of diagrams is very important,
because they determine effective potentials and renormalization functions of
the quantum field models. Moreover, for various physical processes, the regular
parts of diagrams may be evaluated numerically (sometimes with rather
qood quality) but their singular parts should be known analytically because
many types of them should be canceled in the end of calculations.\\

{\bf 2.} To evaluate $A=f(1,1)$ we represent the r.h.s. of Eq.(\ref{3}) as the sum of two terms $A_1$ and $A_2$.

We introduce the new variable $s=(1-x)/2$ and represent $A_2$ as
\begin{eqnarray}
A_2 \equiv -\int_0^1 dx \frac{z\log{z}}{1-z} = -4\int_0^1  \frac{ds}{3+s^2} 
\log{\Bigl(\frac{1-s^2}{4}\Bigr)}
\nonumber
\end{eqnarray}

The term $A_1$ may be rewritten in the form ($y=1-t$)
\begin{eqnarray}
A_1 \equiv \int_0^1 dx  \int^{1-z}_0 dy \frac{\log{(1-y)}}{y}
= \int_0^1 ds \int^{4/(1-s^2)}_1 dt \frac{\log{t}}{1-t}
\label{4}
\end{eqnarray}

Changing the order of integration in the r.h.s. of (\ref{4}), we have
\begin{eqnarray}
A_1  = 6\int_0^1  \frac{ds}{3+s^2} 
\log{\Bigl(\frac{1-s^2}{4}\Bigr)}
\nonumber
\end{eqnarray}

Thus, the evaluation of 
\begin{eqnarray}
A  = 4\int_0^1  \frac{ds}{3+s^2} \biggl[
\log{(1-s)} + \log{(1+s)} - \log{4} \biggr]
\nonumber
\end{eqnarray}
is very
simple. Integrating by part, we obtain the final result
\begin{eqnarray}
A = -\frac{2}{\sqrt{3}} Cl_2\Bigl(\frac{\pi}{3}\Bigr),
\label{5}
\end{eqnarray}
where
$Cl_2(\theta )$ is Clausen integral \cite{Lewin}
\begin{eqnarray}
Cl_2(\theta) = \int_0^{\theta} 
\log{(2sin(\theta '/2))} d \theta ' 
\nonumber
\end{eqnarray}

Repeating above calculations, we have
\begin{eqnarray}
B =
-\frac{4}{\sqrt{3}} Cl_2\Bigl(\frac{\pi}{3}\Bigr)~~~\mbox{ and }~~~
C
= -\frac{1}{2} \log^2{3} + \frac{4}{3\sqrt{3}} Cl_2\Bigl(\frac{\pi}{3}\Bigr)
\label{6}
\end{eqnarray} \\

{\bf 3.} We have obtained results for the singular parts of a FD class 
in closed analytical form. These analytical results are important in the
calculations of physical processes of Standard Model, where many terms
may be involved and the verification and the evaluation of the
singularities is very important problem.
Moreover, the constant A is only one numerical factor in the recent calculation
\cite{ChCh} of three-loop correction to the effective potential of $O(N)$
$\fii^4$ model:
\begin{eqnarray}
-\frac{(4\pi)^6}{\la^2} \cdot V^{(3)}_{eff}(\fii_c) &=&
\frac{\la^2\fii_c^4}{4} \biggl\{\frac{1129}{192} + \frac{A}{8} \biggr\}
+
\frac{m^2\la\fii_c^2}{2} \biggl\{-\frac{25}{96} - \frac{A}{4} \biggr\}
+ \Biggl[
\frac{\la^2\fii_c^4}{4} \biggl\{-\frac{629}{96} - \frac{3A}{4} -\zeta_3\biggr\}
\nonumber \\
&+&
\frac{m^2\la\fii_c^2}{2} \biggl\{-\frac{287}{48}  \biggr\}
+
m^4 \biggl\{\frac{25}{96} + \frac{A}{4} \biggr\} \Biggr]\cdot
\log{\biggl(1+\frac{\la\fii_c^2}{2m^2} \biggr)}
+ \Biggl[
\frac{\la^2\fii_c^4}{4} \cdot \frac{143}{48} \nonumber \\
&+&
\frac{m^2\la\fii_c^2}{2} \cdot \frac{17}{6}
+
m^4  \cdot \frac{11}{48}  \Biggr]\cdot
\log^2{\biggl(1+\frac{\la\fii_c^2}{2m^2} \biggr)}
- \Biggl[
\frac{\la^2\fii_c^4}{4}  \cdot \frac{9}{16}    \nonumber \\
&+&
\frac{m^2\la\fii_c^2}{2}  \cdot \frac{7}{12}
+
m^4 \cdot \frac{5}{48}    \Biggr]\cdot
\log^3{\biggl(1+\frac{\la\fii_c^2}{2m^2} \biggr)}
\nonumber
\end{eqnarray}

After above-mentioned calculation of $A$ (i.e. Eq.(\ref{5})), this three-loop
correction
becomes known fully analytically.\\

I am gratefull to Drs. M.Kalmykov and M.Tentyukov for the interest to this
work and to Dr. J.-M. Chung for the correspondences.\\
The work has been supported in part by Russian Found
for Fundamental Investigations (Grant N 98-02-16923).

\end{document}